\documentclass[journal]{IEEEtran}
\usepackage{amsfonts} 
\usepackage{mathtools}
\usepackage{graphicx}
\usepackage{booktabs}
\usepackage{url}
\usepackage[caption=false,font=footnotesize]{subfig}
\usepackage{authblk}
\usepackage{array}
\usepackage{algpseudocode}

\graphicspath{{figs/}}
\mathtoolsset{showonlyrefs}

\newcommand{\Rad}{\mathcal{R}}
\newcommand{\prox}{\text{prox}}
\newcommand{\argmin}{\text{argmin}}
\newcommand{\code}[1]{\texttt{#1}}
\title{Four-dimensional tomographic reconstruction by time domain decomposition}

\author{Viktor~V.~Nikitin,
        Marcus~Carlsson,
        Fredrik~Andersson,
        and Rajmund~Mokso
\thanks{V.V. Nikitin and R. Mokso are with the MAX IV Laboratory, Lund University, Lund, Sweden, email: (viktor.nikitin@maxiv.lu.se, rajmund.mokso@maxiv.lu.se)}
\thanks{M. Carlsson and F. Andersson are with the Centre for Mathematical Sciences, Lund University, Lund, Sweden, email: (marcus.carlsson@math.lu.se, fredrik.andersson@math.lth.se)}
}
\begin{document}
\maketitle

\begin{abstract}
Since the beginnings of tomography, the requirement that the sample does not change during the acquisition of one tomographic rotation is unchanged.
We derived and successfully implemented a tomographic reconstruction method which relaxes this decades-old requirement of static samples. In the presented method, dynamic tomographic data sets are decomposed in the temporal domain using basis functions and deploying an L1 regularization technique where the penalty factor is taken for spatial and temporal derivatives. We implemented the iterative algorithm for solving the regularization problem on modern GPU systems to demonstrate its practical use.
\end{abstract}

\section{Introduction}
\IEEEPARstart{T}{ime} resolved four-dimensional X-ray computed tomography is widely used in medicine and material sciences where the inner structure of the sample under study is dynamically changing in time.  
The conventional approach to the dynamic tomography is to acquire measurement data during rotation with a constant angular step size. Then, the reconstruction is performed for each $180$ degrees rotation cycle by using reconstruction methods such as Filtered Back-projection (FBP) or Algebraic Reconstruction Technique (ART). The obtained series of recovered objects then form temporal samples representing the object evolution in time. This scheme, however, assumes that the object is static during each $180$ degrees interval. Any significant change in the object's structure during a single rotation cycle introduces motion artifacts apparent as blurred and corrupted reconstructions. There is therefore a constant race between the speed of sample rotation and the sample dynamics. Shorter scan times may be achieved by decreasing the detector exposure time or by reducing the number of projections for each $180$ degrees. However, in practice, the reduced detector exposure time leads to a lower signal to noise ratio, so as the limited number of projections gives specific incompleteness artifacts. 

Algorithms for four-dimensional tomographic reconstruction became of great interest especially with the development of fast detector systems in the last years. With brilliant synchrotron light sources, it is possible to perform continuous data acquisition with more than 8{GB}/s rate \cite{Mokso_JSR2017} and produce terabytes of acquired three-dimensional data sets and corresponding four-dimensional reconstructions in a single experiment. Processing of such a big data requires fast and optimized reconstruction algorithms and powerful software-hardware systems.

Various methods have been proposed for reconstructing tomography data for continuous dynamic acquisition. The first class of methods is based on estimating a prior information about the actual motion. 
For instance, space-time Gibbs priors define relationships among neighboring points in space and time by using information about the motion, see \cite{lalush1998block,lee2005study,gravier2006fully}. The algorithms proposed in \cite{van2017registration,kabus2009evaluation} in turn operate with estimated deformation vector fields (DVFs) between different time frames \cite{chen2008simple}. With a prior information, it is possible to keep reconstruction quality by decreasing the number of projection for a 180 degrees interval. Half-circle intervals representing different time frames can thus be scanned faster, that results in the suppression of motion artifacts. 
In the method proposed in \cite{ruhlandt2017four}, the authors generalize the usual back-projection along straight lines to dynamically curved paths constructed according to the motion model obtained from projections. Dynamically curved paths are then utilized for compensating deformations.

The second class of methods for suppressing motion artifacts takes into account the regularization in a non-local fashion. The methods analyze similarity between corresponding patches at different time steps, even if the patches have moved to another location. In \cite{yang2012robust} the non-local regularization penalty is an unweighted sum of distances between patch pairs in the three-dimensional object. Temporal Non-local Means (TNLM) method proposed in \cite{jia20104d} utilizes weighting factors defined according to the ground truth objects. Another recent research was carried out by Kazantsev et. al \cite{kazantsev20154d,kazantsev2013gpu}. The authors estimate local structural correlations over multiple time frames in order to find the edges inside the object which remain constant in time, then the patched-based regularization (analog of the non-local regularization) is performed according to the obtained object structure. The authors also propose a sparsity seeking approach that operates with a non-local penalty for collecting only relevant information in the spatial and temporal spaces, see \cite{kazantsev2016temporal}. This approach sufficiently decreases the amount of computations.

The third class of methods for the four-dimensional tomographic reconstruction is built upon the concept of compressed sensing. Compressed sensing employs sparsity promoting algorithms with using the $L_1$ norm minimization. The spatial-temporal total variation regularization (STTVR) introduced by Wu et. al \cite{wu2012spatial} operates with $L_1$ penalty factors for the gradient in spatial variables (Total Variation penalty) and for the temporal variable, independently. The method proposed by Ritschl et.al \cite{ritschl2012iterative} extends the Total Variation penalty to the  temporal dimension by taking the gradient in four variables. Corresponding iterative scheme for solving regularization problem consists of two steps. The first step is fidelity term minimization by ART or CGLS, and the second is Total Variation minimization by the gradient-descent method. Authors note that such a separate minimization is caused by data dimensionality and high computational costs. There also exist methods where the penalty factor is represented by $L_p$ norm with $p\in(0,1)$, see \cite{sidky2014constrained,chartrand2013nonconvex}. It is shown that in some cases the norm $L_p$ is more effective than $L_1$ norm because it is closer to $L_0$ - a direct measure of sparsity. 

The regularization problem that arises with employing the concept of compressed sensing for four-dimensional tomographic data reconstruction is non-smooth, therefore standard methods such as least-squares, conjugate gradients are not applied directly.
For most non-smooth problems, a global optimum cannot be found with a given precision and in a reasonable time. The quality of solutions often entirely depends on the model, initial values, and optimization algorithms.
The most common way for solving the regularization problem in four-dimensional tomography is by splitting the minimization function by parts and working with each part independently. Solutions from different parts are combined only after a particular number of iterations. Concerning the compressed sensing in tomography, the minimization is firstly performed for the fidelity term that includes the projection operator, then the contribution from the sparseness terms is taken into account. The primal-dual algorithm proposed by Chambolle et al. \cite{chambolle2011first} allows solving the non-smooth minimization problem directly, i.e. by minimizing the resulting function for all terms together within one iteration. The authors proved that the algorithm has $O(\frac{1}{N})$ convergence rate in finite dimensions, and $O(\frac{1}{N^2})$ if the primal or the dual objective function is uniformly convex. The algorithm is presented as a general framework with established connections to other known algorithms such as Arrow-Hurwicz method \cite{arrow1958studies}, Douglas-Rachford splitting algorithm \cite{eckstein1992douglas,he20121}, and preconditioned ADMM \cite{esser2010general,benning2015preconditioned}. 

An alternative regularization technique successfully used for the static tomography is a sparse data representation by using appropriate functions. In \cite{lee2001wavelet} the function set of wavelets is proposed for decomposition of the projection operator. The reconstruction is then
based on thresholding of the noisy wavelet coefficients. Similar procedure of the tomographic data inversion is performed by using other types of functions: curvelets
\cite{candes2000curvelets}, shearlets \cite{cerejeiras2011inversion}, Gabor frames \cite{colonna2010radon}.

In this work, we use the concept of compressed sensing in the way that data in the temporal direction is represented by a linear combination of appropriate basis functions, and the $L_1$ norm minimization is performed for the gradient in both spatial and temporal variables. The choice of basis functions depends on the motion structure inside the object and can be determined according to the measured data. For solving obtained non-smooth regularization problem, we adopt the primal-dual Chambolle-Pock algorithm \cite{chambolle2011first}.

There are two main advantages of the proposed method compared to the methods mentioned above. First, in contrast to other approaches, we address the cases where rapid motions happen during 180 degrees interval. Other methods operate with different strategies for decreasing the number of projections to represent this interval, and, in that way, require small amplitude of structural changes inside the interval. Second, the proposed method allows operating with big real four-dimensional data that has many samples in the temporal direction. This is achieved by the time-domain data decomposition that sufficiently decreases data sizes. Reconstruction, in this case, are reasonably fast, whereas the existing methods are time-consuming to an extension that impairs their practical use for standard-sized data volumes.

The paper is organized as follows. In Section 2, we introduce the projection operator in 4D, and explain the regularization strategy. Section 3 shows how the minimization problem for recovering  objects can be solved by the Chambolle-Pock algorithm. How to decrease the number of computation by representing data as a linear combination of appropriate basis functions is discussed in Section 4. In section 5 we validate our approach on synthetic data, while in section 6 we process two experimental data sets. Details for technical algorithm acceleration by utilizing high-performance facilities with CPUs and GPUs are given in Section 7. Conclusions and outlook are given in Section 8.

\section{Regularization problem}
Let $f(x,y,z,t)$ be function which represents a three-dimensional object dynamically changing in time $t$. The object is rotated continuously and projection data is measured for angles $\theta$ and for radial direction $s$. The projection operator in this case is described by integration over the lines through the object state at time $t$, which is connected to the rotation angle $\theta$. We assume a linear connection between the angle $\theta$ and time $t$, i.e. $\theta=\alpha t$, where the parameter $\alpha$ in practice is related to the speed of rotation and the detector exposure time. The projection operator $\Rad_\alpha:\mathbb{R}^3\times[0,T]\to \mathbb{R}^2 \times \mathit{S}^1$ where $\mathit{S}^1$ denotes the unit circle, is defined by
\begin{equation}
\begin{aligned}
\Rad_\alpha& f(s,z,\theta)=\\&\iiint f(x,y,z,t)\delta(x\cos\theta\!+\!y\sin\theta\!-\!s)\delta(\theta-\alpha t) dx\,dy\,dt
\end{aligned}\label{Rfwd}
\end{equation}
The corresponding adjoint operator $\Rad_\alpha^*:\mathbb{R}^2\times \mathit{S}^1 \to \mathbb{R}^3\times[0,T]$ is defined as follows
\begin{equation}
\begin{aligned}
\Rad_\alpha^* g(x,y,&z,t)=\\&\iint g (\theta,s,z)\delta(x\cos\theta\!+\!y\sin\theta\!-\!s)\delta(\theta-\alpha t)ds\,d\theta
\end{aligned}\label{Radj}
\end{equation}

The inverse problem of recovering the function $f$ from the measurements $g=\Rad_\alpha f$ has plenty of possible solutions. The non-uniqueness is caused by the fact that at each particular time frame $t$ there exist only one projection related to the angle $\theta=\alpha t$, which is surely not enough to recover the unique object structure. In this case, regularization can be used to introduce assumptions on the solution.
The traditional approach to finding a unique solution is by minimizing the data fidelity term as
\begin{equation}
\begin{aligned}
\hat{f}=\text{argmin}_f\left\{\frac{1}{2}\lVert \Rad_\alpha f-g\rVert_2^2\right\}
\end{aligned}\label{l2}
\end{equation}
This term is commonly used in static tomography where the object does not change during $180$ degrees rotation. Since the cost function is quadratic, one can use gradient-based methods such as the standard least-squares iteration scheme, or the conjugate gradient least-squares scheme with a faster rate of convergence. It is also common to use tomography specific methods. Indeed, recovering the object structure from a limited number of the measured projection angles can be done by algebraic reconstruction methods \cite{andersen1984simultaneous,jiang2003convergence,andersen1989algebraic}, suppressing the Poisson noise in reconstruction is typically done by the EM algorithm \cite{mclachlan2007algorithm,wu1983convergence}.  

None of the listed methods operate with data along the time axis and, consequently, could produce a big number of possible solutions for \eqref{l2}.
Thus, we consider additional penalty factor $J(f)$ and focus on the following regularization problem 
\begin{equation}
\begin{aligned}
\hat{f}=\text{argmin}_f\left\{\frac{1}{2}\lVert \Rad_\alpha f-g\rVert_2^2+\lambda J(f)\right\},
\end{aligned}\label{lr}
\end{equation}
where $\lambda$ is a regularization parameter establishing a trade-off between the data fidelity term and the regularization term.

How to solve the minimization problem \eqref{lr} depends on the structure of the penalty term $J(f)$. The gradient-based algorithms can be used for the quadratic penalty $J(f)=\lVert Kf\rVert_2^2$, where $K$ is a linear operator such as the gradient $\nabla$. With $J(f)=\lVert Kf\rVert_1$ the resulting function becomes non-differentiable and more complicated methods have to be used. The methods assume computation of the proximal operator which is used to make an approximation to a value while making a compromise between the accuracy of the approximation and a cost associated to the new value. There exist proximal splitting methods that are based on the fact that the functions $\frac{1}{2}\lVert \Rad_\alpha f-g\rVert_2^2$ and $\lambda J(f)$ are used individually yielding an easily implementable algorithm. These methods include, for instance, Forward-Backward splitting \cite{combettes2011proximal}, Douglas–Rachford splitting \cite{eckstein1992douglas,he20121}, and Alternating-Direction Method of Multipliers (ADMM) \cite{esser2010general,benning2015preconditioned}. The primal-dual algorithm proposed by Chambolle et al. \cite{chambolle2011first} also operates with proximal operators, however, minimization of the resulting function is performed for all terms together.
Non-smooth penalties such as $J(f)=\lVert Kf\rVert_p$ for $p\in(0,1)$ are beneficial for even more sparse results, cf. \cite{sidky2014constrained,chartrand2013nonconvex}. However, the developed algorithms are more complicated and additional assumptions are required to guarantee convergence.

In this work we introduce the regularization term described with respect to four variables as
\begin{equation}
\begin{aligned}
\lambda_1\big\lVert &|\nabla_{\lambda_2}f|\big\rVert_1=\\&\lambda_1\left\lVert\sqrt{\left(\frac{\partial f}{\partial x}\right)^2+\left(\frac{\partial f}{\partial y}\right)^2+\left(\frac{\partial f}{\partial z}\right)^2+\left(\lambda_2\frac{\partial f}{\partial t}\right)^2}\right\rVert_1,
\end{aligned}\label{TV1}
\end{equation}
where constant $\lambda_1$ denotes a trade-off between data fidelity and  the regularization term, and constant $\lambda_2$ controls the level of sparseness in the temporal direction.
The inclusion of the temporal derivative in the regularization term results in preserving rapid data changes and diminishes small data changes in the temporal direction. This assumes that the object under study does not rapidly change the whole structure, i.e., most object parts keep constant between adjacent time frames. 
The inclusion of derivatives in spatial directions, in turn, preserve sharp edges of the object inner structure and minimize noise components. Moreover, spatial derivatives allow controlling possible artifacts coming from the temporal derivative since they penalize rapid temporal data change of adjacent spatial points. 

As a result, with regularization term \eqref{TV1} we establish a connection between spatial and temporal variables. Alternatively, one can consider spatial and temporal penalty factors independently. However, in this case the temporal sparseness may not be controlled by the object inner structure.
Finally, we propose to use the following model for recovering function $f$ from the measurements $g$

\begin{equation}
\begin{aligned}
\hat{f}=\text{argmin}_f\left\{\frac{1}{2}\lVert \Rad_\alpha f-g\rVert_2^2+\lambda_1\big\lVert |\nabla_{\lambda_2} f|\big\rVert_1 \right\}
\end{aligned}\label{lr12}
\end{equation}
For solving this regularization problem we decided to use the first-order, primal-dual algorithm of
Chambolle and Pock because of its general formulation. In \cite{chambolle2011first} the authors show that most popular algorithms, including Douglas–Rachford splitting and ADMM, are particular cases of the Chambolle-Pock algorithm.

\section{Chambolle-Pock algorithm for dynamic data reconstruction}
In this section, we will recapitulate main structure of the  Chambolle-Pock algorithm and show how to apply it to the regularization scheme \eqref{lr12}. 

The algorithm operate with proximal operators defined for a function $F(f)$ as follows
\begin{equation}
\begin{aligned}
\prox_{\sigma F}(h) = \argmin_f \left\{F(f)+\frac{1}{2\sigma}\lVert h-f\rVert ^2_2\right\},
\end{aligned}\label{prox}
\end{equation}
where $\sigma$ defines a trade-off between two minimization terms. The proximal operator can be interpreted as an approximation to a value, while making a compromise between the accuracy of the approximation $\lVert h-f\rVert ^2_2$ and a cost associated to the new value (function $F(f)$). 
We refer to \cite{parikh2014proximal} for more details about proximal operators and applications.
The algorithm is designed for solving the saddle-point problem described by
\begin{equation}
\begin{aligned}
\hat{f}=\argmin_f \left\{\max_h \big(\langle Kf,h \rangle + G(f) +F^*(h)\big)\right\},
\end{aligned}\label{genform}
\end{equation}
which is the dual formulation of the problem $\hat{f}=\argmin_f \left\{F(Kf)+G(f)\right\}$.
Here $K$ is a linear map, $F^*,G$ are proper, convex, lower-semicontinuous (l.s.c.) functions, $F^*$ being itself the convex conjugate of a convex l.s.c. function $F$. 
The algorithm in Figure \ref{chpock} solves the problem \eqref{genform} with $1/N$ rate of convergence. 

\begin{figure}
\begin{algorithmic}[1]
\State Initialize: $\theta\in[0,1]$, $\tau\sigma\lVert K\rVert^2<1$, $(f_0,h_0)$ is some initial guess, $\tilde{f}^0=f^0$
\Repeat
\State $h^{n+1} = \prox_{\sigma F^*}(h^n+\sigma K\tilde{f}^n)$
\State $f^{n+1} = \prox_{\tau G}(f^n-\tau K^*h^{n+1})$
\State $\tilde{f}^{n+1} = f^{n+1}-\theta(f^n-f^{n+1})$
\Until convergence criteria met
\end{algorithmic}
\caption{Chambolle-Pock algorithm. General formulation.}
\label{chpock}
\end{figure}
The algorithm can be adapted for solving \eqref{lr12} with $F(Kf)=F_1(K_1f)+F_2(K_2f)$ ($G(f) = 0$), where linear operator $K_1$ acting on $f$ produce functions $p_1$ which then are used as input for functional $F_1$,
\begin{align*}
&p_1=K_1 f = \Rad_\alpha{f}, \qquad F_1(p_1) = \frac{1}{2}\lVert p_1 - g\rVert_2^2, 
\end{align*}
and where linear operator $K_2$ acting on $f$ produces a spatial vector field $\vec{p}_2$ that is used as an input for functional $F_2$,
\begin{align*}
\vec{p}_2=K_2 f = \nabla_{\lambda_2} f, \qquad F_2(\vec{p}_2)=\lambda_2\big\lVert |\vec{p}_2| \big\rVert_1.
\end{align*}
The corresponding dual problem is obtained by adding three dual variables $(h_1,\vec{h}_2)$ with respect to $(p_1,\vec{p}_2)$.  It has the following form
\begin{equation}
\begin{aligned}
\hat{f}\!=\!\argmin_f\!\max_{h_1,\vec{h}_2}\!\left\{\langle K_1f,h_1 \rangle\!+\!\langle K_2f,\vec{h}_2 \rangle\!+\!F_1^*(h_1)\!+\!F_2^*(\vec{h}_2)\!\right\}
\end{aligned}
\end{equation}
Corresponding convex conjugate functions for $F_1,F_2$ and adjoint operators for $K_1,K_2$ are computed as follows
\begin{equation}
\begin{aligned}
& K_1^*(p_1) = \Rad_\alpha^* p_1,\qquad K_2^*(\vec{p}_2) = -\text{div}_{\alpha_2}\,\vec{p}_2, \\
& F_1^*(h_1) = \max_{h_1'} \{\langle h_1,h_1'\rangle - \frac{1}{2}\lVert h_1'-g\rVert_2^2\}=\langle h_1,g\rangle+\frac{\lVert h_1\rVert_2^2}{2},\\
& F_2^*(\vec{h}_2) = \max_{\vec{h}_2'} \{\langle \vec{h}_2,\vec{h}_2'\rangle - \lambda_1 \big\lVert |\vec{h}_2'|\big\rVert_1 \}=\kappa(\vec{h}_2),
\end{aligned}
\end{equation}
where $-\text{div}_{\lambda_2}$ is the adjoint operator to $-\nabla_{\lambda_2}$, and $\kappa(\vec{h}_2)=0$ if  $\lambda\big\lVert |\vec{h}_2| \big\rVert_1\le 1$, otherwise $\kappa(\vec{h}_2)=\infty$. Now it is straightforward to compute corresponding proximal operators:

\begin{flalign}
\prox_{\sigma\!F_1^*}(h_1)\!=\! \argmin_{h_1'}\!\big\{\langle h_1',g\rangle+&\frac{\lVert h_1'\rVert_2^2}{2}\! +\!\frac{1}{2\sigma}\lVert h_1\!-\!h_1'\rVert ^2_2\big\}&\\ & =  (h_1-\sigma g)/(1+\sigma),
\end{flalign}
\vspace{-\baselineskip}
\begin{flalign}
\prox_{\sigma\!F_2^*}(\vec{h}_2)\!=\!\argmin_{\vec{h}_2'}\!\big\{\kappa(\vec{h}_2')+&\frac{1}{2\sigma}\lVert \vec{h}_2\!-\!\vec{h}_2'\rVert ^2_2\big\}&\\ & =\vec{h}_2/\max(1,\big\lVert
|\vec{h}_2|\big\rVert_1/\lambda_1),
\end{flalign}
\vspace{-\baselineskip}
\begin{flalign}
&\prox_{\tau\!G}(f)\!=\!\argmin_{f'} \big\{0 +\frac{1}{2\tau}\lVert f\!-\!f'\rVert ^2_2\big\} = f.&
\end{flalign}

We result in the algorithm shown in Figure \ref{chpockadapt} that summarizes main steps for solving \eqref{lr12}.
\begin{figure}
\begin{algorithmic}[1]
\State Initialize: $\theta\in[0,1]$, $\tau\sigma<1^*$, $f^0,h_1^0,\vec{h}_2^0$ is some initial guess, $\tilde{f}^0=f^0$
\Repeat \hfill //description
\State $h^{n+1}_1 = \frac{ h^n_1+\sigma \Rad_\alpha\tilde{f}^n-\sigma g}{1+\sigma}$, \begin{flushright}\small//$h_1^{n+1} = \prox_{\sigma F_1^*}(h_1^n+\sigma K_1\tilde{f}^n)$\end{flushright}\normalsize 
\State $\vec{h}^{n+1}_2 = \frac{\vec{h}^n_2+\sigma \nabla_{\lambda_2}\tilde{f}^n}{\max\left(1,\left\lVert |\vec{h}^n_2+\sigma \nabla_{\lambda_2}\tilde{f}^n|\right\rVert_1/\lambda_1\right)}$,\begin{flushright}\small //$\vec{h}_2^{n+1} = \prox_{\sigma F_2^*}(\vec{h}_2^n+\sigma K_2\tilde{f}^n)$\end{flushright}\normalsize 
\State $f^{n+1} = f^n-\tau \Rad_\alpha^* h^{n+1}_1+\tau\text{div}_{\lambda_2}\,\vec{h}^{n+1}_2,$ \begin{flushright}//\small$f^{n+1} = \prox_{\tau G}(f^n-\tau K_1^*h_1^{n+1}-\tau K_2^*\vec{h}_2^{n+1}))$\normalsize \end{flushright}
\State $\tilde{f}^{n+1} = f^{n+1}+\theta(f^{n+1}-f^n).$
\Until convergence criteria met
\end{algorithmic}
* linear operators $\nabla_{\lambda_2}$, $\Rad_\alpha$ are normalized after a constant multiplication.
\caption{Chambolle-Pock algorithm for dynamic tomography.}
\label{chpockadapt}
\end{figure}

\section{Problem optimization}\label{ch_opt}
The regularization problem \eqref{lr12} becomes computation and memory intensive if one aims to work with the object represented by several hundred or thousands of samples in each spatial and temporal variables. The most computationally intensive parts in the algorithm involve evaluating the projection operator $\Rad_\alpha$ and the corresponding adjoint operator $\Rad_\alpha^*$ operator many times. If we assume that the number of samples in each spatial and in the temporal (or angular) direction is of the order of $N$, then the complexity of computing the operators $\Rad_\alpha$ and $\Rad_\alpha^*$ by directly discretizing the integrals in formulas \eqref{Rfwd} and \eqref{Radj} is $\mathcal{O}(N^4)$. In static tomography there exist several approaches how to decrease computational complexity to $\mathcal{O}(N^3\log N)$, where all angles are used for reconstruction simultaneously. The methods include Fourier-based methods \cite{beylkin1995fast,fessler2003nonuniform}, the log-polar-based method \cite{andersson2005fast,andersson2016fast}, hierarchical decomposition \cite{basu2000n,george2007fast}. These methods are not directly used in the dynamic tomography problem since each projection is related to a particular object state in time. However, in what follows we will show that the methods are still in use for dynamic tomography. 

One way of how to reduce resources for computations is by introducing an additional assumption on the object movement. Let us assume the motion at each concrete space sample $(x,y,z)$ can be approximated by a linear combination of basis functions $\left\{\varphi_j(t)\right\}_{j=0}^{M-1}$, i.e.,
\begin{equation}
\begin{aligned}
f(x,y,z,t)\approx\sum_{j=0}^{M-1}f_j(x,y,z) \varphi_j(t),
\end{aligned}\label{repr}
\end{equation}
where $\left\{f_j(x,y,z)\right\}_{j=0}^{M-1}$ are decomposition coefficients. Choice of basis functions $\varphi_j$ for better approximation depends on the motion structure. As a straightforward example, one can choose the Fourier basis with a low number of coefficients to represent slow motions, and a high number of coefficients to represent rapid motions. Other possible functions for representation include Haar wavelets, Heaviside step functions, as well as their smooth approximations (Gram-Schmidt Orthonormalization \cite{van1996matrix} can be used to make an orthonormal basis if necessary). It should be also noted that typically the object needs to be reconstructed only with some particular step in time so that the total number of time frames is sufficiently smaller than the total number of angles. This fact potentially allows decreasing the number of coefficients for representation.

Now by making use of representation \eqref{repr} and exploiting the linearity property of the projection operator we have

\begin{equation}
\begin{aligned}
\Rad_\alpha f(s,z,\theta)=
\sum_{j=0}^{M-1} \Rad f_j(\theta,s,z) \varphi_j\left(\frac{\theta}{\alpha}\right),
\end{aligned}\label{radapr}
\end{equation}
where the operator $\Rad f(\theta,s,z)=\iint f(x,y,z)\delta(x\cos\theta\!+\!y\sin\theta\!-\!s)dx\,dy$ denotes the standard Radon transform computed for the whole set of angles $\theta$.
For the constructed algorithm (Figure \ref{chpockadapt}) we also need the representation of the adjoint operator $\Rad_\alpha^*$. It can be found through the adjoint equality
\begin{equation}
\begin{aligned}
\langle \Rad_\alpha f,g\rangle\approx&\left\langle \sum_{j=0}^{M-1} \Rad f_j(\theta,s,z) \varphi_j\left(\frac{\theta}{\alpha}\right),g(\theta,s,z)\right\rangle=\\
&\sum_{j=0}^{M-1} \left\langle f_j(\theta,s,z),\Rad^*\left(g\hat{\varphi}_j\right)(x,y,z)\right\rangle=\\
&\left\langle f,\sum_{j=0}^{M-1}\varphi_j(t)\Rad^*\left(g\hat{\varphi}_j\right)(x,y,z)\right\rangle,
\end{aligned}
\end{equation}
which gives the following approximation for the adjoint operator
\begin{equation}
\begin{aligned}
\Rad_\alpha^* g(x,y,z,t)\approx\sum_{j=0}^{M-1}\varphi_j(t)\Rad^*\left(g\hat{\varphi}_j\right)(x,y,z)
\end{aligned}\label{radadjapr}
\end{equation}

Computing the projection operator $\Rad_\alpha$ and its adjoint $\Rad_\alpha^*$ with respect to formulas \eqref{radapr} and \eqref{radadjapr} can sufficiently decrease the number of computations and allocated memory because of two facts. First, as mentioned above, the number of needed coefficients $M$ is typically much smaller than the number of samples in time. Second, the Radon transform  for each coefficient, $\Rad f_j$ in formula \eqref{radapr}, as well as the adjoint operator $\Rad^*(g\hat{\varphi}_j)$ in formula \eqref{radadjapr}, are computed independently on the temporal variable. With this fact, $\Rad f_j$ becomes periodic with a period $2\pi$. Also, the Radon transform for the interval $[\pi,2\pi)$ is the same as the Radon transform for the interval $[0,\pi)$ after changing the sign of variable $s$. So instead of computing the Radon transform  $\Rad f_j$ for the whole set of angles $\theta$ it is enough to compute the transform only for the interval $[0,\pi)$ and simply distribute the result to other angles. In what follows we give a detailed description of how it works, as well as how this idea is adapted for computing $\Rad^*(g\hat{\varphi}_j)$. 

For simplicity we can assume that the object rotation includes exactly $N_\pi$ angular intervals of size $\pi$, i.e. $\theta\in [0,\pi N_\pi)$. In other cases, one can subtract this interval from the whole set of angles, and work with the remaining part separately. For the chosen interval we have $\theta=k\pi+\theta_0$, where $\theta_0\in[0,\pi)$ and $k$ is an integer value. Then we have the following expression for the Radon transform $\Rad f_j (\theta,s,z)$ from formula \eqref{radapr},
\begin{equation}
\begin{aligned}
\Rad f_j(\theta,s,z) = \Rad f_j(k\pi+\theta_0,s,z) = \Rad f_j(\theta_0,(-1)^k s,z)
\end{aligned}
\end{equation}
In turn, the adjoint operator $\Rad^*(g\hat{\varphi}_j)$ from formula \eqref{radadjapr} can be rewritten as follows
\begin{flalign}
&\Rad^*\left( g(\theta,s,z)\hat{\varphi}_j(\theta,s,z)\right)=&\\&\sum_{k=0}^{N_\pi-1}\Rad^*\left( g(k\pi+\theta_0,(-1)^k s,z)\hat{\varphi}_j(k\pi+\theta_0,(-1)^k s,z)\right)=&\\&\Rad^*\left(\sum_{k=0}^{N_\pi-1}g(k\pi+\theta_0,(-1)^k s,z)\hat{\varphi}_j(k\pi+\theta_0,(-1)^k s,z)\right)=\\&\Rad^*\left(\tilde{g}_j(\theta_0,s,z)\right)&
\end{flalign}

where $\tilde{g}_j(\theta_0,s,z)=\sum_{k=0}^{N_\pi-1}g(k\pi+\theta_0,(-1)^k s,z)\hat{\varphi}_j(k\pi+\theta_0,(-1)^k s,z)$. Here instead of computing the adjoint Radon transform for the whole set of angles it is enough to sum up data over angular intervals of size $\pi$ and then compute the adjoint operator only for the angular interval $[0,\pi)$.  

It is straightforward to adapt steps of the algorithm in Figure \eqref{chpockadapt} with corresponding decompositions (\ref{radapr}-\ref{radadjapr}) for discrete spatial variables, discretization in temporal variable in turn needs to be explained. In what follows for simplicity we ignore discretization with respect to spatial variables $x,y,z$ by introducing the following simplified notation
\begin{equation}
\begin{aligned}
	&\mathsf{f}(t)=f(x,y,z,t),\\
    &\mathsf{R}\mathsf{f}(\theta)=\Rad f(\theta,s,z).
\end{aligned}
\end{equation}
Now let us assume that the total number of samples in the angular direction is $N_\theta$, and the total number of reconstructed time frames is $N_t$, where in practice we are interested in $N_t\ll N_\theta$. For instance, the first time frame is associated with projection angle $0^\circ$, the second - to $180^\circ$, the third - to $360^\circ$, etc.  
We let $m$ be the amount of angles per time frame, i.e. we connect samples in the temporal direction $\{t\}_{k=0}^{N_t-1}$ to the angular samples $\{\theta_k\}_{k=0}^{N_\theta-1}$ in the following way
\begin{equation}
\begin{aligned}
&\theta_k=k\Delta_\theta ,\quad k=0\dots N_\theta-1\\
&t_k=\frac{\theta_{km}}{\alpha}=\frac{km\Delta_\theta }{\alpha}, \quad k=0,\dots N_t-1
\end{aligned}\label{discrt}
\end{equation}
where the parameter $\alpha$ shows a linear connection between temporal and angular samples as in the definition \eqref{Rfwd} of the projection operator. 
With discretization \eqref{discrt} the derivative operator $\frac{\partial f}{\partial t}$ in the algorithm from Figure \ref{chpockadapt} at discrete samples reads as
\begin{equation}
\begin{aligned}
\frac{\mathsf{f}(t_{k+1})-\mathsf{f}(t_k)}{\Delta t}=\frac{\alpha(\mathsf{f}(\theta_{km+m})-\mathsf{f}(\theta_{km}))}{\Delta_\theta}, \,\, k\!=\!0\dots N_t\!-\!1
\end{aligned}
\end{equation}
The discrete version of the projection operator $\Rad_\alpha$ having the approximation \eqref{radapr} is written as
\begin{equation}
\begin{aligned}
\mathsf{R}_\alpha\mathsf{f}(\theta_k)=\sum_{j=0}^{M-1} \mathsf{R} \mathsf{f}_j(\theta_k) \varphi_j\left(\frac{\theta_k}{\alpha}\right), \,\, k\!=\!0\dots N_\theta\!-\!1
\end{aligned}
\end{equation}
where the coefficients $\mathsf{f}_j$ are computed as follows
\begin{equation}
\begin{aligned}
 \mathsf{f}_j=\sum_{k=0}^{N_t-1} \mathsf{f}(t_k)\varphi_j(t_k)=\sum_{k=0}^{N_t-1}\mathsf{f}\left(\frac{\theta_{km}}{\alpha}\right)&\varphi_j\left(\frac{\theta_{km}}{\alpha}\right),\\& \,\, j\!=\!0,\dots M\!-\!1
\end{aligned}
\end{equation}

\section{Validation}
In this section we validate our approach through simulations.

For the validation we use a synthetic model considering only two spatial variables $x,y$ and consisting of circles with varying spatial distribution in time (see Figure \ref{fig:recbub}). The displacement of the spheres is captured in 8 intervals of each consisting by angular views across $\pi$. The two upper panels of the left column show two sequential time frames 3 and 4. The plots also contain circle border marks of the next object state. For instance, the circle border masks of time frame 3 correspond to circle borders of the time frame 4. Note that the displacement of circles between time frames 3 and 4 is smaller than between time frames 4 and 5. 
The model has sizes $(N,N,N_t)=(256,256,8)$, and the projection data has sizes $(N_\theta,N)=(128\times 8,256)=(1024,256)$, where $128$ is the number of projections for covering the interval $[0,\pi)$.
Two upper panels of the middle column in Figure \ref{fig:recbub} show reconstruction by using standard FBP method where the Shepp-Logan filter is utilized for suppressing high frequencies in projections. Reconstruction of time frames 3,4 is performed by using projections from the intervals $[3\pi,4\pi)$ and $[4\pi,5\pi)$, respectively. Motion artifacts in the recovered time frame 3 are relatively small since the object inner structure has not rapid changes between time frames 3 and 4, whereas the recovered time frame 4 has clearly visible motion artifacts due to fast movement of circles between time frames 4 and 5. The same behavior demonstrates the regularized algebraic reconstruction (two upper panels in the right column of Figure \ref{fig:recbub}) where the penalty factor is taken only for spatial derivatives in $x$ and $y$, without any penalty for the temporal derivative. This method suppressed the noise in data and diminished limited angles artifacts, however, it does not deal with motion artifacts.

The bottom six panels in Figure \ref{fig:recbub} demonstrate results of the proposed method, with a different number of basis function used for data representation in the temporal direction. In this example, we consider the Fourier basis with $M=8, 16, 32$ elements defined by
\begin{equation}
\begin{aligned}
\varphi_j(t)=e^{2\pi i t (j-M/2)}, \quad j=0\dots M-1.
\end{aligned}
\end{equation}
Recall that the Fourier basis with a higher number of coefficients is used to recover more rapid motions.
The results show that some motion artifacts are suppressed with $M=8$ basis functions, and almost all artifacts are suppressed for the cases $M=32$ basis functions. 
Selecting the number of basis functions is the main free parameter that the user have to adjust in the proposed method. 

\newcolumntype{C}{>{\centering\arraybackslash}m{0.28\textwidth}}
\begin{figure*}
\begin{tabular}{l*3{C}@{}}
&\small{Synthetic data} & \small{FBP reconstruction} & \small{Iterative, no penalty on $\frac{\partial}{\partial t}$} \\ 
\rotatebox[origin=c]{90}{\small{Time frame 3 of 8}} & 
\includegraphics[width=0.29\textwidth]{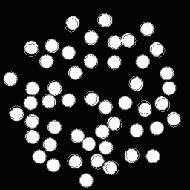} & 
\includegraphics[width=0.29\textwidth]{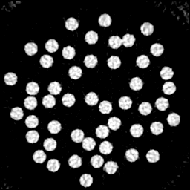} & 
\includegraphics[width=0.29\textwidth]{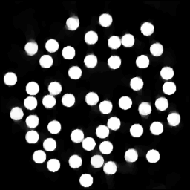} \\ 
\rotatebox[origin=c]{90}{\small{Time frame 4 of 8}} & 
\includegraphics[width=0.29\textwidth]{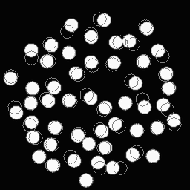} & 
\includegraphics[width=0.29\textwidth]{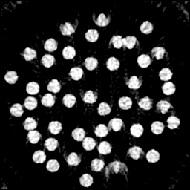} & 
\includegraphics[width=0.29\textwidth]{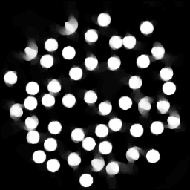} \\ 
& \small{Iterative, basis size M=8} & \small{Iterative, basis size M=16}&  \small{Iterative, basis size M=32}  \\ 
\rotatebox[origin=c]{90}{\small{Time frame 3 of 8}} & 
\includegraphics[width=0.29\textwidth]{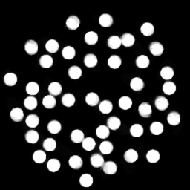} & 
\includegraphics[width=0.29\textwidth]{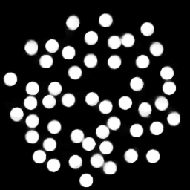} & 
\includegraphics[width=0.29\textwidth]{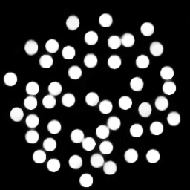} \\ 
\rotatebox[origin=c]{90}{\small{Time frame 4 of 8}} & 
\includegraphics[width=0.29\textwidth]{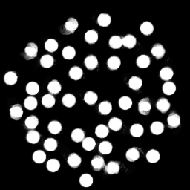} & 
\includegraphics[width=0.29\textwidth]{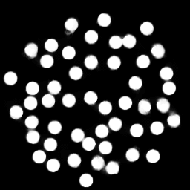} & 
\includegraphics[width=0.29\textwidth]{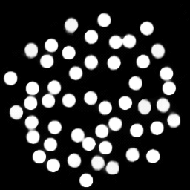} \\ 
\end{tabular}
\caption{Two time frames of the synthetic data and corresponding reconstruction by FBP method and by using the iterative scheme with penalty on the gradient in spatial variables (two top rows), reconstruction by the proposed iterative scheme with penalty on the gradient in spatial and temporal variables, for $M=8,16,32$ basis functions (two bottom rows).}\label{fig:recbub}
\end{figure*}

\section{Applications to experimental data}

In the following we show the reconstruction of experimental data. We investigate the rheology of liquid foams by fast synchrotron X-ray tomographic microscopy \cite{Raufaste_EPL2015}. Foams are complex cellular systems which require artifact free tomographic reconstruction for a reliable quantification of their time-dependent properties such as deformation fields of bubbles. In our example we acquire X-ray projections of the liquid foam flowing through a constriction and being rotated around the tomographic axis at a rate of 840 deg/s. Each X-ray exposure takes 0.7 ms and in total we acquire 130 tomographic data sets. 

The experiment was performed at the TOMCAT beamline of the Swiss Light Source using the fast acquisition setup \cite{Mokso_JSR2017}. The data size is $(N,N_\theta,N_z)=(2016,300\times 130,1800)$, where $300$ is the number of projection inside the interval $[0,\pi)$, and $130$ is the total number of time frames.
Prior to tomographic reconstruction the raw projections were by phase retrieval algorithm to account for the interference occuring due to the partial coherence properties of the synchrotron beam \cite{paganin_2002,mokso_Jph2013}. As an example of reconstruction we considered the data that corresponds to vertical slices $z\in[1200,1328)$.  Two time frames recovered by using static FBP method are shown in Figure \ref{fig:recfoam}, left. The tomogram at time frame 94 does not contain motion artifacts because the amplitude of the sample motion during the acquisition of this time-frame is small enough to meet the static sample assumption. The tomogram at time frame 95 shows motion artifacts due to the rapid movement of bubbles between time frames 95 and 96. As seen in Figure \ref{fig:recfoam}, right, by applying our new method (algorithm in Figure \ref{chpockadapt}) we overcome the motion artifacts in time frame 95, but also maintain the image quality in the static case (frame 94). For reconstruction we use $M=16$ basis functions, parameters $\lambda_1,\lambda_2$ showing trade-off between data fidelity and the derivative parts (see formulation \eqref{lr12}) have been experimentally chosen as $2^{-12}$ and $2^2$, respectively. The iterative scheme in the algorithm is performed for 512 iterations, the result of reconstruction by the FBP is chosen as an initial guess.

We have also tested our method for reconstructing dynamic tomography data recently acquired at 2-BM beamline at Advanced Photon Source. The sample was prepared as the slurry of ceramic particles in alcohol. Along with alcohol evaporation, the ceramic particles aggregated to form clusters.  
In total, 12 tomographic data sets were continuously measured with 6 deg/s rotation rate and 30 ms exposure time.
The data size is $(N,N_\theta,N_z)=(2560,900\times 12,700)$, where $900$ is the number of projection inside the interval $[0,\pi)$. The left part of Figure \ref{fig:recslurry} shows two time frames corresponding to angles $\theta\in[7\pi, 8\pi]$ and vertical slices $z\in(250,314)$ recovered by using static FBP method. Both reconstructed time frames contain motions artifacts that are mostly seen at the regions where ceramic particles are moving in the direction of the vertical central axis. The proposed method substantially suppress motion artifacts due to the rapid movement of particles, see Figure \ref{fig:recslurry}, right. It should be noted, that the method also improves the resolution level since the structure of particles becomes more reliable for segmentation. For reconstruction we use $M=24$ basis functions, parameters $\lambda_1,\lambda_2$ have been experimentally chosen as $2^{-9}$ and $2^2$, respectively. The iterative scheme in the algorithm from Figure \ref{chpockadapt} is performed for 512 iterations, the result of reconstruction by FBP is chosen as an initial guess. Note the noise reduction effect of the regularized reconstruction as compared to FBP. This is more apparent in this dataset than in the case of the foam data because the phase retrieval used in the foam example acts similar to a low pass filter. 

\newcolumntype{D}{>{\centering\arraybackslash}m{0.44\textwidth}}
\begin{figure*}
\begin{tabular}{l*2{D}@{}}
& FBP reconstruction & Iterative, basis size M=16\\ 
\\
\rotatebox[origin=c]{90}{Time frame 94 of 130} & 
\includegraphics[width=0.45\textwidth]{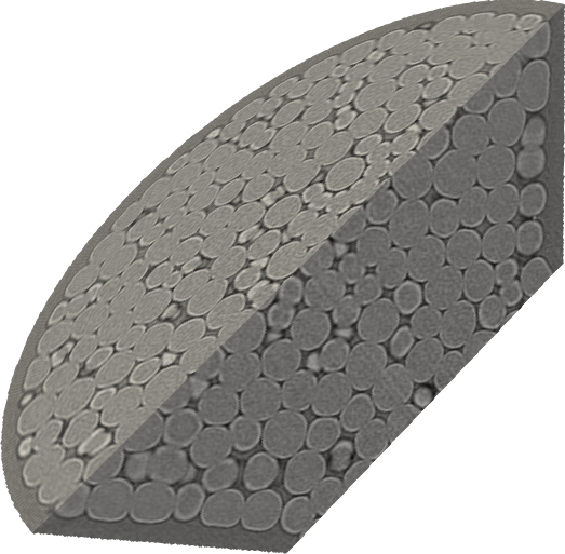} & 
\includegraphics[width=0.45\textwidth]{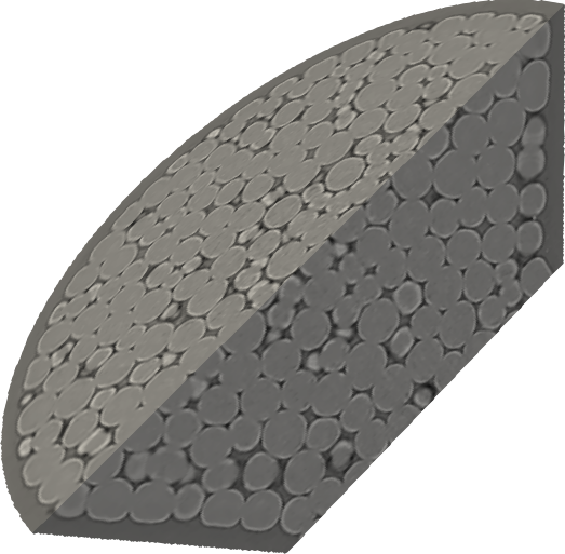} \\ 
\\
\rotatebox[origin=c]{90}{Time frame 95 of 130} & 
\includegraphics[width=0.45\textwidth]{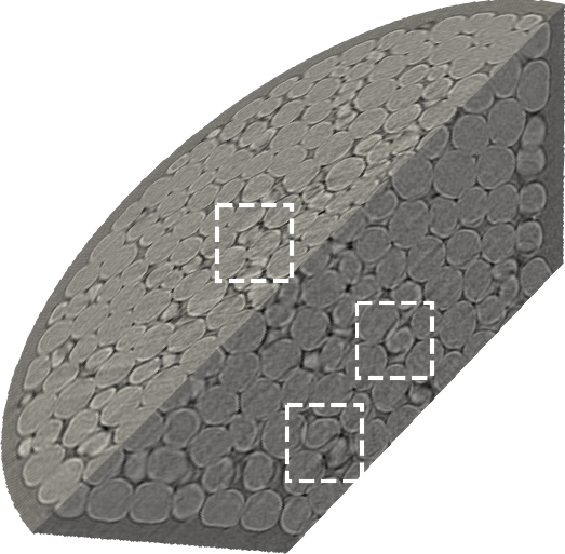} & 
\includegraphics[width=0.45\textwidth]{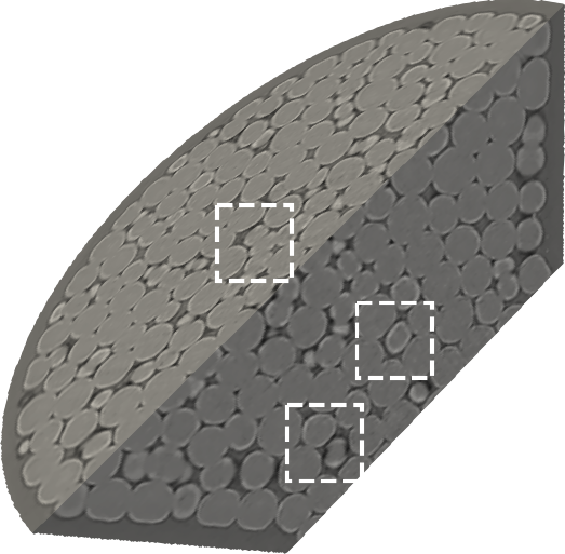} \\ 
\end{tabular}
\caption{Two time frames of the foam in a glass (half-cropped to show vertical slices) reconstructed by FBP method (left) and  by the proposed iterative scheme with penalty on the gradient in spatial and temporal variables (right).}\label{fig:recfoam}
\end{figure*} 

\begin{figure*}
\begin{tabular}{l*2{D}@{}}
& FBP reconstruction & Iterative, basis size M=24\\ 
\\
\rotatebox[origin=c]{90}{Time frame 7 of 12} & 
\includegraphics[width=0.45\textwidth]{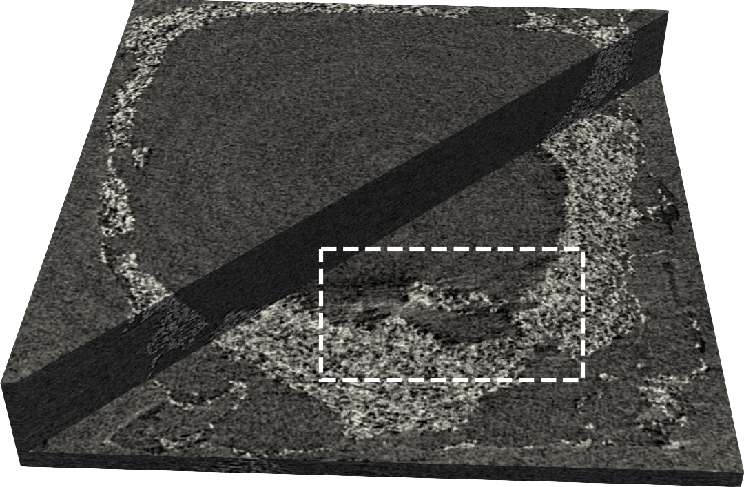} & 
\includegraphics[width=0.45\textwidth]{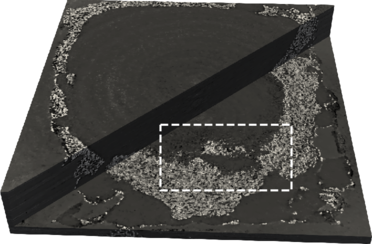} \\ 
\\
\rotatebox[origin=c]{90}{Scaled part} & 
\includegraphics[width=0.45\textwidth]{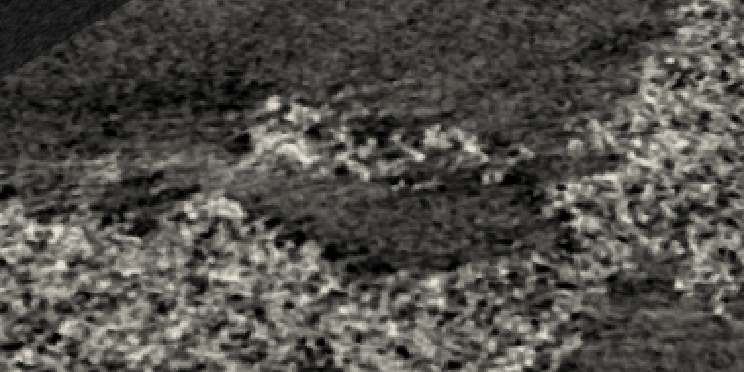} & 
\includegraphics[width=0.45\textwidth]{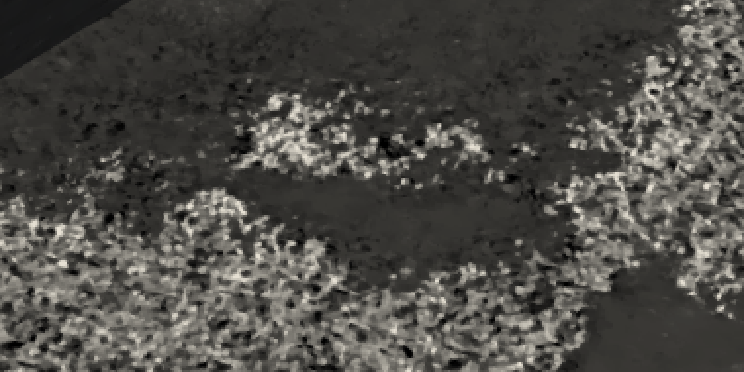} \\
\\
\rotatebox[origin=c]{90}{Time frame 8 of 12} & 
\includegraphics[width=0.45\textwidth]{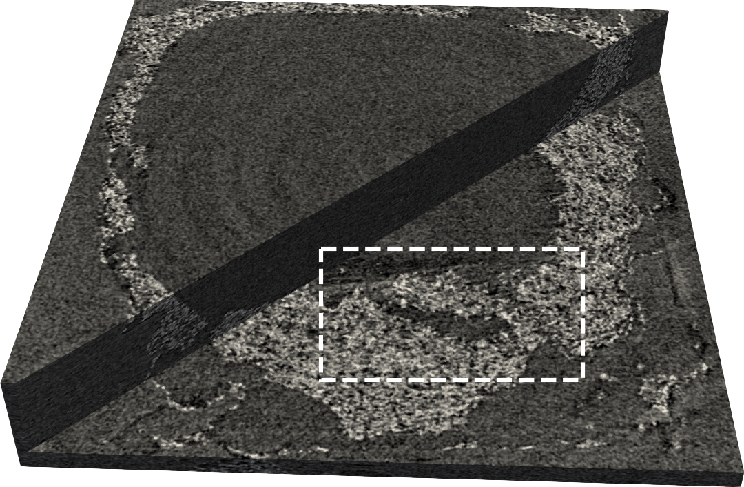} & 
\includegraphics[width=0.45\textwidth]{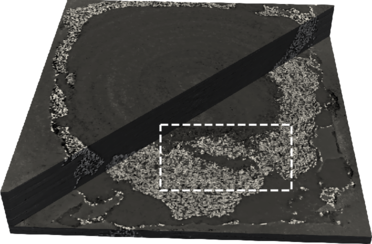} \\ 
\\
\rotatebox[origin=c]{90}{Scaled part} & 
\includegraphics[width=0.45\textwidth]{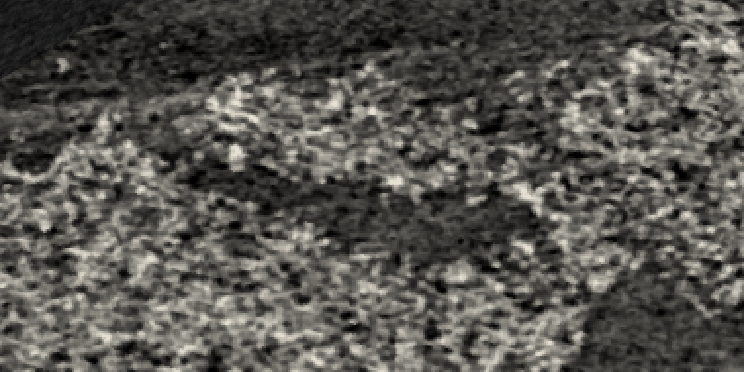} & 
\includegraphics[width=0.45\textwidth]{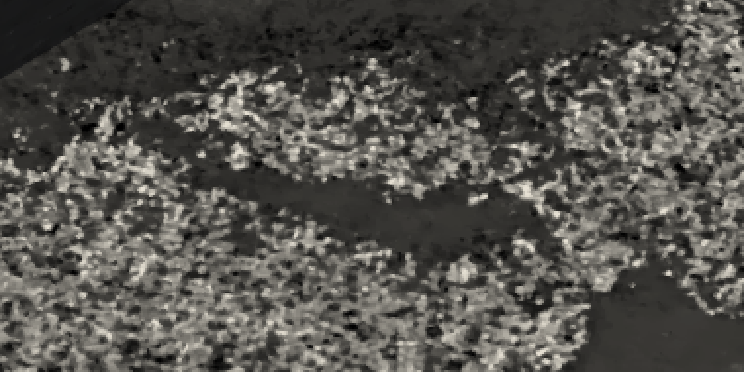} \\ 
\end{tabular}
\caption{Two time frames of ceramic particles aggregation in alcohol, reconstructed by FBP method (left) and  by the proposed iterative scheme with penalty on the gradient in spatial and temporal variables (right). }\label{fig:recslurry}
\end{figure*}

\section{Parallel implementation}
The reconstruction problem \eqref{lr12} remains resource intensive even after considering decomposition techniques described in Section \ref{ch_opt}. Besides the mathematical improvement of the algorithm, we propose its technical acceleration by using parallel computing on CPUs and GPUs. Parallel computations on CPUs are typically done with making use of OpenMP and MPI technologies. NVidia CUDA technology in turn provides an interface to accelerate computations on GPUs. The main difference between these two platforms is that the number of computational cores on GPUs is more than 100 times bigger than the number of cores on CPU. However, GPU cores are much slower than CPU cores, moreover, the memory handling mechanism on GPU is not so advanced as the mechanism on CPU. 
Nevertheless, GPUs have shown themselves as a powerful tool to accelerate algorithms in tomography \cite{andersson2016fast,scherl2007fast,gac2008high,van2016fast}. This is because most reconstruction algorithms contain a lot of small identical operations that can be computed independently. We will give some details of how the algorithm from Figure \ref{chpockadapt} is parallelized for computing on several GPUs, and compare its parallel implementation to the OpenMP implementation on CPU.

The most resource intensive parts of the algorithm are the Radon transform $\Rad_\alpha$ in Step 3, and the corresponding adjoint operator $\Rad_\alpha^*$ in Step 5. We have already described in Section \ref{ch_opt} how these operators are computed through the standard Radon transform and the back-projection for the angular interval of size $\pi$, see formulas \eqref{radapr} and \eqref{radadjapr}. There exist several fast GPU implementations for computing these standard operators. For the proposed algorithm we decided to use the log-polar-based method \cite{andersson2005fast} implemented on GPU since it demonstrates the best performance compared to known analogs, cf. \cite{andersson2016fast}. We have also constructed fast GPU kernels for evaluation of gradient and divergence operators in Steps 4 and 5, as well as for evaluation of standard algebraic operations in Steps 3-6.

The time-resolved reconstruction algorithm must operate with four-dimensional big data sets, therefore, computations have to be split by parts so that the processing data fit into the operating memory, and to the memory on GPU.
A straightforward approach generally used for reconstructing tomography data is to split computations by slices (in the $z$-variable) where all slices are recovered independently. Typically one GPU is used to recover a set of slices that fit into the GPU memory. This approach has to be slightly modified in our case since the algorithm also includes computations of derivatives in the $z$ direction. To deal with that we transfer to GPU two extra slices located before and after the current set of slices processed by one GPU. There are no thread concurrency errors since the additional data from the two slices is kept read-only.
Suppose that we intend to parallelize the algorithm by using \code{Nth} number of threads denoted by \code{ith=0 : Nth-1}. For simplicity, we assume that the total number of slices is a multiple of \code{Nth}. The procedure is done for Steps 3..6 located inside the iterative loop \textbf{repeat}..\textbf{until}, synchronization of parallel threads is done after Step 6. Parallel computation of Step 3 is done by separating the whole set of slices by parts associated with the thread number \code{ith} as 
\code{z = ith*Nz/Nth : (ith+1)*Nz/Nth-1}. 
Steps 4 and 5 include computation of operators $\nabla_{\lambda_2}$ and $\text{div}_{\lambda_2}$ that assumes using extra slices in the $z$ direction. Therefore, in Step 3 for the input read-only function $\tilde{f}^n$ we use slices with numbers \code{z = max(0,ith*Nz/Nth-1) : min((ith+1)*Nz/Nth,Nz-1)} and compute the vector field $\vec{h}_2$ for \code{z = max(0,ith*Nz/Nth-1) : (ith+1)*Nz/Nth-1}. In turn, in Step 4 the new function $f^{n+1}$ appears for \code{z = ith*Nz/Nth : (ith+1)*Nz/Nth-1} by making use of the input vector field $\vec{h}_2$. Parallel computation of the Step 6 is straightforward with \code{z = 
ith*Nz/Nth-1 : (ith+1)*Nz/Nth-1}. Switching to the next iteration $n\to n+1$ is performed after thread synchronization.

The GPU implementation is also optimized by making use of the technique for overlapping data transfers with computations. Nvidia CUDA library provides the streams technology for simultaneous execution of different code parts. One optimal solution is to create three streams for 1) Host to Device data migration, 2) Computation on GPU, 3)  Device to Host data migration. While one data part is computed, the next part can be loaded to GPU memory, as well as the result of the previous part can be unloaded from GPU memory. This strategy is especially good with introducing NVLink technology \cite{foley2017ultra} that allows fast bidirectional copy between CPU and GPU.

We note that the parallelization strategy proposed above can be also used to accelerate computations by using OpenMP and MPI technologies. In this case, the whole dataset is split by parts in order to fit the operating memory. Overlap of computations and data transfers between operating memory and hard disks is done by different parallel OpenMP threads.

Performance tests are carried out by making use of the foam data from the previous section. Recall that, the foam data with sizes $(N_\theta,N,N_z)=(130\cdot 300,2016,1800)$ produces the object of sizes $(N,N,N_z,N_t)=(2016,2016,1800,130)$. There is typically no need to recover all time frames by using the proposed algorithm since only particular time frames contain motion artifacts. Thus, for the performance test, we decrease the total number of time frames to a smaller value, $N_t=8$. The linear binning procedure \cite{handrick2006evaluation} with reducing data sizes is used to demonstrate scalability and computational complexity of the algorithm. Bins of sizes 4,2, and 1 (no binning) in each spatial direction produce data sizes $(N,N_z)=(504,450),(N,N_z)=(1008,900),(N,N_z)=(2016,1800)$, respectively. Number of basis functions for the function representation \eqref{repr} in the temporal direction is chosen as $M=16$ since it demonstrates appropriate quality for reconstruction, see Figure \ref{fig:recfoam}. All computations are carried out in single precision. Table \ref{tab:times} shows average time for one iteration of the algorithm in Figure \ref{chpockadapt} for different platforms. Graphical processors Tesla P100 and Tesla V100 demonstrate 7-11 performance gain compared to Intel Xeon E5-2650 processor with 12 cores and 24 threads with hyper-threading. Tesla P100 has a fast NVLink connection between CPU and GPU. This type of connection allows bidirectional data transfers with more than 3 times faster speed compared to PCI Express 3.0. Such a fast connection gives almost linear performance gain when utilizing a system with 4 Tesla P100 connected to one CPU, see the last column in Table \ref{tab:times}. Computational times for the system with 4 Tesla V100 connected to one CPU with PCI Express 3.0 are not presented in the table due to the lack of time scaling caused by high system bus load.

The last observation from Table \ref{tab:times} is that computational complexity of the algorithm corresponds to the complexity $\mathcal{O}(N_z N^2 \log N)$ of computing the Radon transform with the log-polar-based method. For GPU systems one can observe that 2x increase of the binning size gives approximately 8x decrease in computational times (7-9x in reality due to hardware specific reasons). 

\setlength\extrarowheight{2pt}
\begin{table*}\centering
	\caption{Average time (in seconds) for 1 iteration of the proposed algorithm (Figure \ref{chpockadapt}) for reconstruction $N_t$ time frames of sizes $(N,N,N_z)$ from the measurements of sizes $(N,N_\theta,N_z)$ by using $M$ basis functions. 
Case: $\,N_\theta=2400, \,M=16,\,N_t=8, \,(N,N_z)$ are chosen with respect to 4,2,1 binning.}
    \begin{tabular}{ | c| c | c | c | c|}
		\hline
		 $N,N_z$ (binning) &  \shortstack{Intel Xeon E5-2650\\(12 cores/24 threads)} & \shortstack{Tesla V100 \\ (PCI Express 3.0)} & \shortstack{Tesla P100 \\ (NVLink)} & \shortstack{4x Tesla P100 \\ (NVLink)}  \\
\hline 
504, 450 (4) & 1.5e+02 & 1.2e+01 & 2.1e+01 & 6.2e+00 \\ 
\hline 
1008, 900 (2) & 1.1e+03 & 9.6e+01 & 1.5e+02 & 4.8e+01 \\ 
\hline 
2016, 1800 (1) & 7.7e+03 & 8.7e+02 & 1.2e+03 & 4.2e+02 \\ 
\hline 
	\end{tabular}\label{tab:times}
\end{table*}

\section{Conclusions and outlook}
We derived, validated and applied a new method for reconstruction of four-dimensional tomographic data sets by time-domain decomposition and this way for the first time couple directly space and time domains.  Our approach works on continuous acquisition where multiple time frames are recorded through multiple tomographic rotations of $\pi$ while sample motion is not restricted within the individual intervals.  Motion artifacts have been fully suppressed in by selecting the corresponding number of basis functions. The implementation on modern GPU systems demonstrates 7-11 performance gain compared to one modern CPU with 12 cores and 24 threads. Computational times results are acceptable for their use in practice.

The source code is publicly available (\url{https://github.com/math-vrn/rectv_gpu}). The foam data with an example script for reconstruction by the proposed method can be downloaded via Tomobank \cite{de2018tomobank}, see section Datasets - Dynamic - Foam data.   

We assume that the proposed method can be improved in terms of performance. First, the adapted Chambolle-Pock algorithm has $\mathcal{O}(1/N)$ rate of convergence, where $N$ is the iteration number. In our opinion, the convergence rate may probably be increased to $\mathcal{O}(1/N^2)$ with some assumptions on the data structure, that will sufficiently accelerate the whole reconstruction process. Second, different types of representation basis, such as Haar wavelets or Heaviside step functions, may be considered in order to decrease the number of decomposition coefficients, and thus increase the performance. Finally, the parallel implementation on several GPUs may potentially be faster with utilizing new technologies provided by latest versions of NVidia CUDA library. We are going to address all these problems in our further research.

\section{Acknowledgments}
This work was funded by the Crafoord Foundation (2016-0691) and by the
Swedish Research Council grant (2017-00583). We acknowledge the Paul Scherrer Institut, Villigen, Switzerland for provision of synchrotron radiation beamtime (20150185) at the beamline TOMCAT of the Swiss Light Source. The foam samples were prepared by C. Raufaste, B. Dollet and S. Santucci.
Data for ceramic particles aggregation in alcohol was acquired by using resources of the Advanced Photon Source, a U.S. Department of Energy (DOE) Office of Science User Facility operated for the DOE Office of Science by Argonne National Laboratory under Contract No. DE-AC02-06CH11357. We acknowledge Xianghui Xiao from 2-BM beamline for providing this data. 
\bibliographystyle{IEEEtran} 
\bibliography{refs}
\end{document}